# Interaction-free all-optical switching at low light Intensities in a multi-atom cavity QED system

Bichen Zou[1], Zheng Tan[2], Mohamed Musa[3], and Yifu Zhu[1]

[1]Department of Physics, Florida International University, Miami, Florida 33199
[2]Wuhan Institute of Physics and Mathematics, Chinese Academy of Sciences, China
[3]Department of Mathematics and Natural Sciences, Gulf University for Science and Technology, Kuwait

**Abstract:** We propose a scheme of the interaction-free all-optical switching in a multi-atom cavity QED system consisting of three-level atoms confined in a cavity and coupled by a free-space control laser. A signal laser field is coupled into the cavity and its transmission through and reflection from the cavity form two output channels. We show that the free-space control laser induces destructive quantum interference in the excitation of the intra-cavity normal mode, which can be used to switch on or off of the output signal light of the reflection and the transmission channels. When the control light is coupled to the atoms, the input signal light is nearly totally reflected and this means that there is no direct coupling of the control light and the signal light through the cavity-confined atoms. We present analytical and numerical calculations that show this type of the interaction-free all-optical switching in the coupled cavity-atom system can be realized with high switching efficiencies and at low light intensities.

1. Introduction

   Optical devices based on light-controlling-light are important for applications in optical communications and quantum information networks. One aspect of light-controlling-light is all-optical switching, in which a signal light field is switched on or off by a control light field which has attracted a lot of attention in recent years [1-16]. All-optical switching relies on the large optical nonlinearities induced in an optical medium by the coupling of the signal and control lights. The switching efficiency is limited by the achievable optical nonlinearities and the photon loss due to absorption or scattering. In order to operate an all-optical switch at low light levels down to single photons, it is necessary to generate large optical nonlinearities and eliminate the photon loss in an optical medium coupled by the signal light and the control light. Recently, schemes of interaction-free all-optical switching have been proposed and analyzed [17-19]. In these schemes, the direct coupling of the signal light to the control light is eliminated and it is shown that it is possible to suppress the signal photon loss from the absorbing medium. Several experimental demonstrations of the interaction-free all-optical switching based on the 2$^{nd}$ order $\chi^{(2)}$ optical nonlinearities have been reported under conditions of moderate to high light intensities [20-21].

   Here we propose a scheme of the interaction-free all optical switching that is based on the coherent quantum interference effect and can be operated at low light intensities. The model system consists of multiple three-level atoms confined in an optical cavity. A weak signal light is coupled into one of the two normal modes of the multi-atom cavity QED system and a weak control light is coupled to the cavity-confined atoms from free-space. The control light is tuned to the resonant frequency of the normal mode and induces the destructive interference that suppresses the normal mode excitation from the signal light [22-24]. We show that by turning on and off of the control light, the reflected signal light from the cavity and the transmitted signal light through the cavity are switched on or off. Thus the cavity-atom system can be used to realize all-optical switching of the signal light by the control light with two output channels: the reflected signal channel and the transmitted signal channel. Specifically, when the control light is off, the normal mode is excited and the transmitted light is at the maximum (the transmission channel is closed) and the reflected light is suppressed (the reflection channel is open); when the control light is turned on, the transmitted light is suppressed (the transmission channel is open) and the reflected light is at the maximum (the reflection channel is closed). The reflection channel and the transmission channel are complementary and offer versatile applications. When the control light is present, the destructive interference is induced and the intra-cavity light is suppressed, so the input signal light is prevented from coupling into the cavity. Correspondingly, there is no direct coupling between the control light and the signal light, fulfilling the requirement of the interaction-free criterion. Since the destructive interference suppresses the intra-cavity light absorption, the all-optical switching in the cavity-atom system can be realized at low control light intensities for ultra-weak signal light.

We studied the transmission channel for the application of the cavity-atom system in all-optical switching and cross-phase modulation at low light intensities in earlier reports [22-24]. Here we show that the cavity-atom system is versatile and can be used as a two-channel switch with reflection and transmission as complementary channels. We point out the important interaction-free characteristics of the system and derive full analytical results that quantify the all-optical switching performance. Our analysis indicates that under normal operating conditions, the reflection channel has higher switching efficiencies than the transmission channel. Furthermore, we quantify the frequency bandwidth and switching time of the interaction-free all-optical switching in the cavity-atom system and present calculations of the time evolution of the reflection and transmission channels of an input signal pulse.

We note that the proposed scheme for the interaction-free all-optical switching is different from the scheme based on the cavity electromagnetically induced transparency (EIT) [25-28], in which either cavity-confined three level atoms coupled by two laser fields (the signal and the control) or cavity confined four-level atoms coupled by three laser fields are used as the nonlinear media [10, 29-31]. Even though our scheme uses a configuration similar to the cavity EIT scheme based on three-level atoms coupled by two laser fields, there are essential differences between the two schemes. In the cavity EIT scheme, the signal field and the control field are both resonant with the respective atomic transitions ($\Delta_p = \Delta = 0$). When the control field (at $\Delta = 0$) is present, the signal field (at $\Delta_p = 0$) is transmitted through the cavity while the signal reflection from the cavity is minimized [26]. Both the signal field and the control field interact simultaneously with the atoms and are directly coupled. Therefore, the scheme based on the three-level cavity EIT system is not interaction free.

In the following, we derive analytical results and present numerical calculations that characterize the performance of the all-optical switching in the multi-atom cavity QED system and quantify the efficiency of the all-optical switching versus the system parameters such the ground state decoherence, the control light intensity, and the number of atoms in the cavity mode. Then we present an analysis of the frequency bandwidth and switching time of the coupled cavity-atom system.

## 2. Theoretical model and equations

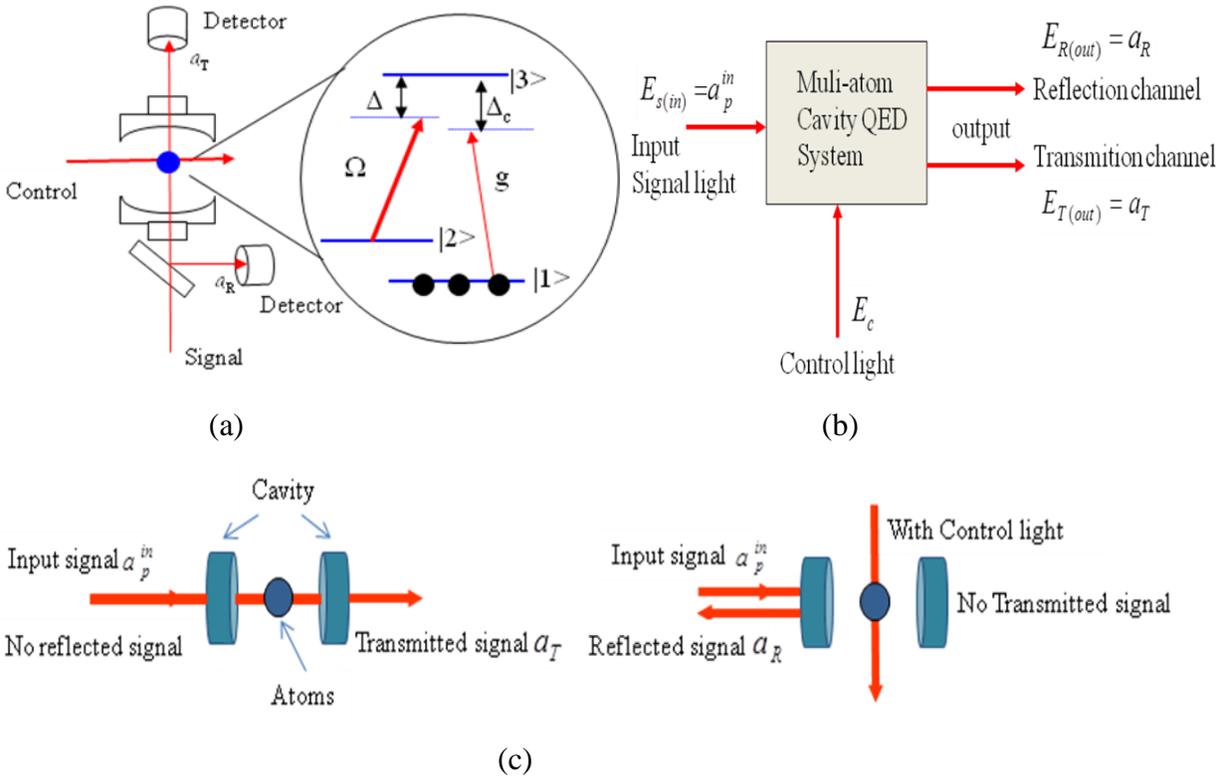

Fig.1 (a) Schematic diagram of the interaction-free all-optical switching in a coupled atom-cavity system. A signal laser is coupled into the cavity, and the transmitted signal light and the cavity reflected signal light are collected by two detectors. A control laser is coupled to the cavity-confined atoms from free space and controls the states of the transmitted signal light and the reflected signal light. (b) Schematic diagram of the light input and output channels. (c) Outcome of the all-optical switching operation. When there is no control light, the transmission channel is closed and the reflection channel is open; when the control light is present, the transmission channel is open and the reflection channel is closed. During the switch operation, there is no direct coupling between the control light and the input signal light.

Fig. 1(a) shows the schematic diagram for the all-optical switching based on a composite atom-cavity system that consists of a single mode cavity containing N Λ-type three-level atoms interacting with a control laser from free space. The cavity mode couples the atomic transition |1>-|3> and the classical control laser drives the atomic transition |2>-|3> with Rabi frequency 2Ω. $\Delta = \nu - \nu_{23}$ is the control frequency detuning and $\Delta_c = \nu_c - \nu_{13}$ is the cavity-atom detuning. A signal laser is coupled into the cavity mode and its frequency is detuned from the atomic transition |1>-|3> by $\Delta_p = \nu_p - \nu_{13}$. Fig. 1(b) shows the schematic diagram of the input-output channels. The input signal light is coupled into the cavity and results in two output channels: the reflected signal light from the cavity and the transmitted signal light though the cavity. The states of the two output channels are controlled by the free-space control light. The detailed output states are depicted in Fig. 1(c). It will be shown that under ideal conditions, when the control light is absent, there is output signal light from the transmission channel, but no light from the reflection channel; when the control light is on, the signal light cannot be coupled into the cavity, the output is switched to the reflection channel and there is no output light from the transmission channel. Therefore, the signal light and the control light do not interact with the atoms simultaneously and the interaction-free requirement is satisfied.

The interaction Hamiltonian for the coupled cavity-atom system is

$$H = -\hbar(\sum_{i=1}^{N} \Omega \hat{\sigma}_{32}^{(i)} + \sum_{i=1}^{N} g \hat{a} \hat{\sigma}_{31}^{(i)}) + H.C. , \qquad (1)$$

where $\hat{\sigma}_{lm}^{(i)}$ (l, m=1-3) is the atomic operator for the ith atom, $g = \mu_{13}\sqrt{\omega_c / 2\hbar\varepsilon_0 V}$ is the cavity-atom coupling coefficient, and $\hat{a}$ is the annihilation operator of the cavity photons. Assuming the equal coupling strength and uniform light fields for the N identical atoms inside the cavity, the resulting equations of motion for $\hat{\sigma}_{lm}^{(i)} = \hat{\sigma}_{lm}$ and the intra-cavity light field (two-sided cavity, one input) are given by [32]

$$\dot{\sigma}_{11} = \gamma_{31}\sigma_{33} + iga\sigma_{31} - iga^+\sigma_{13} \qquad (2\text{-}1)$$

$$\dot{\sigma}_{22} = \gamma_{32}\sigma_{33} + i\Omega\sigma_{32} - i\Omega^*\sigma_{23} \qquad (2\text{-}2)$$

$$\dot{\sigma}_{33} = -(\gamma_{31} + \gamma_{32})\sigma_{33} - iga\sigma_{31} + iga^+\sigma_{13} - i\Omega\sigma_{32} + i\Omega^*\sigma_{23} \qquad (2\text{-}3)$$

$$\dot{\sigma}_{12} = -(\gamma_{21} + i(\Delta_p - \Delta))\sigma_{12} - i\Omega^*\sigma_{13} + iga\sigma_{32} \qquad (2\text{-}4)$$

$$\dot{\sigma}_{13} = -(\frac{\gamma_{31}+\gamma_{32}}{2} + i\Delta_p)\sigma_{13} + iga(\sigma_{33}-\sigma_{11}) - i\Omega\sigma_{12} \qquad (2\text{-}5)$$

$$\dot{\sigma}_{23} = -(\frac{\gamma_{31}+\gamma_{32}}{2} + i\Delta)\sigma_{23} + i\Omega(\sigma_{33}-\sigma_{22}) - iga\sigma_{21} \qquad (2\text{-}6)$$

$$\dot{a} = -((\kappa_1+\kappa_2)/2 + i(\Delta_c-\Delta_p))a - igN\sigma_{31} + \sqrt{\kappa_1/\tau}\,a_p^{in} \qquad (2\text{-}7)$$

where $a_p^{in}$ is the input signal field, $\kappa_i = \frac{T_i}{\tau}$ (i=1-2) is the loss rate of the cavity field on the mirror i ($T_i$ is the mirror transmission and $\tau$ is the photon round trip time inside the cavity). We consider $\gamma_{31} = \gamma_{32} = \Gamma_3$ ($\Gamma_3$ is the decay rate of the excited state $|3\rangle$) and a symmetric cavity such that $\kappa_1 = \kappa_2 = \kappa$. $\gamma_{12}$ is the decoherence rate between the ground states $|1\rangle$ and $|2\rangle$. We are interested in the parameter regime near the normal mode resonance in which the laser fields are near or at resonance with the normal mode transitions, and under the conditions of low light intensities in which the intra-cavity field is very weak and the control field is below the saturation level. It then can be shown that a weak control light induces the destructive quantum interference in the normal mode excitation [22], which can be used to control the amplitude of the cavity transmitted and the cavity reflected signal field. The primary control parameters of the cavity-atom system for the interaction-free all-optical switching are the frequency and intensity of the control laser that are characterized by the control detuning $\Delta$ and control Rabi frequency $\Omega$ respectively.

## 3. Results

We consider the situations where the cavity is tuned on resonance with the atomic transition ($\Delta_c = 0$) and $g \ll \Omega$ such that the atomic population is concentrated in $|1\rangle$. Then the steady-state solution of the intra-cavity probe field is given by

$$a = \frac{\sqrt{\kappa/\tau}\,a_p^{in}}{\kappa - i(\Delta_c - \Delta_p) - i\chi}, \qquad (3)$$

where $\chi$ is the atomic susceptibility given by $\chi = \dfrac{-ig^2 N}{\Gamma_3 - i\Delta_p + \dfrac{\Omega^2}{\gamma_{12} - i(\Delta_p - \Delta)}}$. The reflected signal field from the cavity is given by $a_R = \sqrt{\kappa\tau}\,a - a_p^{in}$ and the transmitted signal field through the

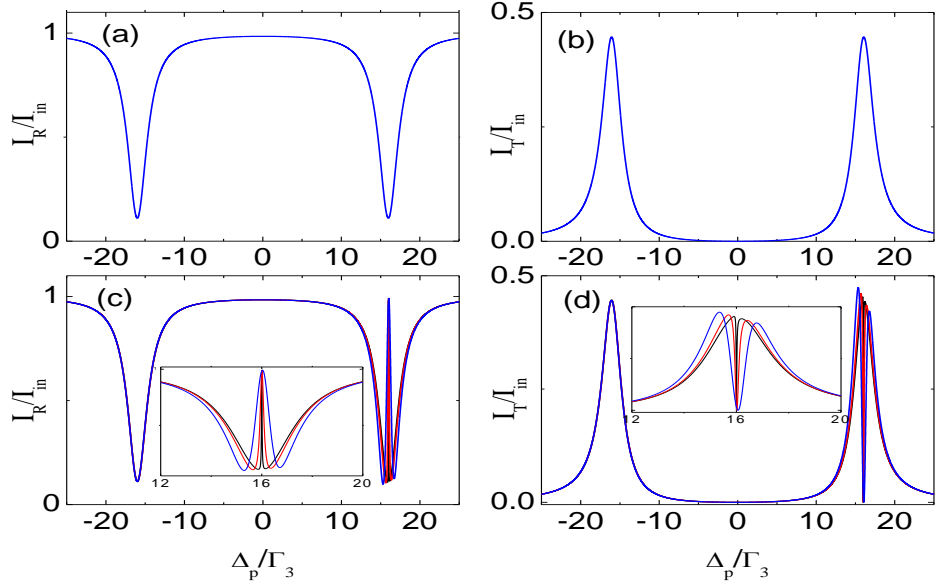

Fig. 2 (a) The reflected signal intensity ($I_R/I_{in}$) and (b) the transmitted signal intensity ($I_T/I_{in}$) versus the signal frequency detuning $\Delta_p/\Gamma_3$ when the control laser is absent ($\Omega=0$). (c) $I_R/I_{in}$ and (d) $I_T/I_{in}$ versus $\Delta_p/\Gamma_3$ when the control laser is present and its frequency is tuned to $\Delta = g\sqrt{N}$ (black line: $\Omega=0.2\Gamma_3$; red line: $\Omega=0.5\Gamma_3$ and blue line: $\Omega=\Gamma_3$). The inset figures show the expanded view across the normal mode resonance $\Delta_p = g\sqrt{N} = 16\Gamma_3$. The other parameters are $\kappa = 2\Gamma_3$, $\gamma_{12}=0.001\Gamma_3$, and $\Delta_c = 0$.

cavity is given by $a_T = \sqrt{\kappa\tau}a$. In order to show that the control field induces the destructive interference in the normal mode excitation of the intra-cavity signal light, we first present the calculated reflection spectra and the transmission spectra of the signal light versus the signal frequency detuning without and with the control laser field. Fig. 2(a) plots the reflected signal light intensity $I_R = a_p a_p^*$ normalized by the input signal intensity $I_{in} = a_p^{in}(a_p^{in})^*$ and Fig. 2(b) plots the transmitted signal light intensity $I_T = a_T a_T^*$ normalized by $I_{in}$ versus the normalized signal laser detuning $\Delta_p/\Gamma_3$ without the control laser ($\Omega=0$). The two peaks in the transmitted signal spectrum represent the two normal modes separated in frequency by the vacuum Rabi frequency $2g\sqrt{N}$ [32-35]. Concomitantly, the reflected signal light exhibits two dips at the resonance of the two normal modes. Fig. 2(c) and Fig. 2(d) plot the reflection spectra and transmission spectra when the control laser is present (($\Omega\neq0$) and its frequency is tuned to the normal mode resonance ($\Delta = g\sqrt{N}$). The spectra show that at the normal mode resonance ($\Delta_p = g\sqrt{N}$), the transmitted signal is suppressed while the reflected signal light is maximized. The narrow spectral peak for the reflected signal light and the narrow spectra dip for the transmitted signal light show that the control laser field induces the destructive interference for the normal mode excitation at $\Delta_p = g\sqrt{N}$ from the intra-cavity signal laser. The signal light transmission through the cavity is suppressed, and the signal light is essentially totally reflected from the cavity at $\Delta_p = g\sqrt{N}$. Therefore, the control laser can be used to turn on or turn off the signal light transmission and the signal light reflection, and the coupled atom-cavity system can be used to perform all-optical switching on a weak signal light by a weak control light. When the control laser is coupled to the cavity-confined atoms, the signal light cannot be coupled into the cavity and is therefore reflected from the cavity. The all-optical switching is performed without the direct coupling between the control light and the signal light and is therefore interaction free.

Next we derive analytical results for the reflected signal field and transmitted signal field, and analyze the interaction-free all-optical switching characteristics of the cavity-atom system. Under the normal mode resonance condition for both the signal laser and the control laser ($\Delta = \Delta_p = g\sqrt{N}$), the reflected signal field is

$$a_R = \frac{ig\sqrt{N}(\gamma_{21}\Gamma_3 + \Omega^2)a_p^{in}}{\kappa(\Omega^2 + \gamma_{21}\Gamma_3) - ig\sqrt{N}(\gamma_{21}\kappa + \gamma_{21}\Gamma_3 + \Omega^2)}. \tag{4}$$

The intensity of the reflected signal light is then

$$I_R = \frac{g^2 N(\Omega^2 + \gamma_{21}\Gamma_3)^2 I_{in}}{\kappa^2(\Omega^2 + \gamma_{21}\Gamma_3)^2 + g^2 N(\gamma_{21}\kappa + \gamma_{21}\Gamma_3 + \Omega^2)^2}. \tag{5}$$

The performance of an optical switch can be characterized by the switching efficiency defined as $\eta = \frac{I(|1\rangle) - I(|0\rangle)}{I_{in}}$, where $I(|1\rangle)$ is the signal output intensity when the switch is closed and $I(|0\rangle)$ is the signal output intensity (leakage) when the switch is open ($I(|1\rangle) > I(|0\rangle)$ and $I_{in}$ is the input signal intensity). Eq. (5) indicates that $I_R(\Omega \neq 0) > I_R(\Omega = 0)$. So for the all-optical switching operating on the reflection output channel, we designate $I_R(\Omega = 0) = I(|0\rangle)$ for the open state $|0\rangle$ of the switch and $I_R(\Omega \neq 0) = I(|1\rangle)$ for the closed state $|1\rangle$ of the switch, then the switching efficiency for the reflection output channel is

$$\eta_R = \frac{I_R(|1>)-I_R(|0>)}{I_{in}} = \frac{g^2 N(\Omega^2+\gamma_{21}\Gamma_3)^2}{\kappa^2(\Omega^2+\gamma_{21}\Gamma_3)^2+g^2 N(\gamma_{21}\kappa+\gamma_{21}\Gamma_3+\Omega^2)^2} - \frac{g^2 N\Gamma_3^2}{\kappa^2\Gamma_3^2+g^2 N(\kappa+\Gamma_3)^2}. \quad (6)$$

For an ideal switch, the switching efficiency η=1. Assuming a strong collective atom-cavity coupling ($g\sqrt{N} \gg \kappa$, $\Omega$, and $\Gamma_3$), the reflected signal intensity is $I_R(|0>) = I_R(\Omega=0) \approx \frac{\Gamma_3^2 I_{in}}{(\kappa+\Gamma_3)^2}$ and $I_R(|1>) = I_R(\Omega \neq 0) \approx \frac{(\Omega^2+\gamma_{21}\Gamma_3)^2 I_{in}}{(\gamma_{21}\kappa+\gamma_{21}\Gamma_3+\Omega^2)^2}$. $I_R(|1>)$ decreases as $\gamma_{21}$ increases so the switching efficiency decreases with increasing $\gamma_{21}$. Usually $\gamma_{21}$ is small ($\gamma_{21} \ll \kappa$, and $\Gamma_3$). If it can be neglected ($\gamma_{21}=0$), then the switching efficiency becomes $\eta_R = \frac{\kappa^2+2\kappa\Gamma_3}{(\kappa+\Gamma_3)^2}$, which indicates that the efficient operation of the all-optical switch in the reflection mode requires a cavity with a decay rate $\kappa > \Gamma_3$.

Next we discuss the all-optical switching with the transmission output channel. The signal light field transmitted through the cavity is derived as

$$a_T = \frac{\kappa(\Omega^2+\gamma_{21}\Gamma_3-ig\sqrt{N}\gamma_{21})a_p^{in}}{\kappa(\Omega^2+\gamma_{21}\Gamma_3)-ig\sqrt{N}(\gamma_{21}\kappa+\gamma_{21}\Gamma_3+\Omega^2)}. \quad (7)$$

The intensity of the transmitted probe field is then

$$I_T = \frac{\kappa^2\{(\Omega^2+\gamma_{21}\Gamma_3)^2+\gamma_{21}^2 g^2 N\}I_{in}}{\kappa^2(\Omega^2+\gamma_{21}\Gamma_3)^2+g^2 N(\gamma_{21}\kappa+\gamma_{21}\Gamma_3+\Omega^2)^2}. \quad (8)$$

Eq. (8) indicates that $I_T(\Omega=0) > I_T(\Omega \neq 0)$. Therefore, for the all-optical switching with the transmission output channel, we designate $I_T(\Omega \neq 0) = I_T(|0>)$ for the open state $|0>$ of the switch and $I_R(\Omega=0) = I_T(|1>)$ for the closed state $|1>$ of the switch. Then, the switching efficiency for the transmission output channel is

$$\eta_T = \frac{\kappa^2 \gamma_{21}^2(\Gamma_3^2+g^2 N)}{\kappa^2(\gamma_{21}\Gamma_3)^2+g^2 N(\gamma_{21}\kappa+\gamma_{21}\Gamma_3)^2} - \frac{\kappa^2\{(\Omega^2+\gamma_{21}\Gamma_3)^2+\gamma_{21}^2 g^2 N\}}{\kappa^2(\Omega^2+\gamma_{21}\Gamma_3)^2+g^2 N(\gamma_{21}\kappa+\gamma_{21}\Gamma_3+\Omega^2)^2}. \quad (9)$$

Assuming strong collective atom-cavity coupling ($g\sqrt{N} \gg \kappa$ and $\Gamma_3$) and neglecting the atomic decoherence ($\gamma_{21}=0$), $I_T(|1>) = I_T(\Omega=0) = \frac{\kappa^2 I_{in}}{(\kappa+\Gamma_3)^2}$ and

$I_T(|0>) = I_T(\Omega \neq 0) \approx \dfrac{\kappa^2 I_{in}}{\kappa^2 + g^2 N} \approx 0$. Then the switching efficiency for the transmission output channel becomes $\eta_T = \dfrac{\kappa^2}{(\kappa+\Gamma_3)^2}$, which again indicates that the efficiency operation of the all-optical switch in the transmission output channel requires a cavity with a decay rate $\kappa > \Gamma_3$. This condition is compatible with the condition for the highly efficient operation of the reflection output channel.

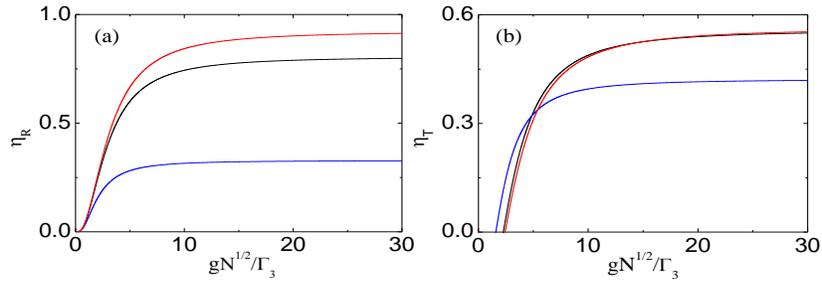

Fig. 3 (a) Switching efficiency of the reflected signal light $\eta_R$ and (b) switching efficiency of the transmitted signal light $\eta_T$ versus $g\sqrt{N}/\Gamma_3$ with $\gamma_{12}=0.0001\Gamma_3$ (red lines), $\gamma_{12}=0.001\Gamma_3$ (black lines) and $\gamma_{12}=0.01\Gamma_3$ (blues lines), respectively. Other parameters are $\Delta p=\Delta= g\sqrt{N}$. $\Omega=0.2\Gamma_3$, and $\kappa=3\Gamma_3$,

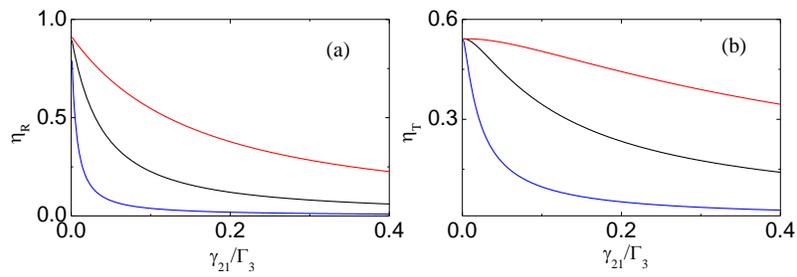

Fig. 4 (a) Switching efficiency of the reflected signal light $\eta_R$ and (b) switching efficiency of the transmitted signal light $\eta_T$ versus the decoherence rate $\gamma_{21}/\Gamma_3$ with $\Omega=\Gamma_3$ (red lines), $\Omega=0.5\Gamma_3$ (black lines) and $\Omega=0.2\Gamma_3$ (blues lines), respectively. Other parameters are $\Delta p=\Delta= g\sqrt{N} =20\Gamma_3$, and $\kappa=3\Gamma_3$,

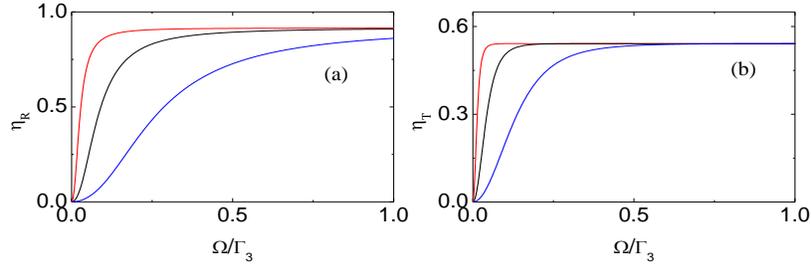

Fig. 5 (a) Switching efficiency of the reflected signal light $\eta_R$ and (b) switching efficiency of the transmitted signal light $\eta_T$ versus $\Omega/\Gamma_3$ with $\gamma_{12}=0.0001\Gamma_3$ (red lines), $\gamma_{12}=0.001\Gamma_3$ (black lines) and $\gamma_{12}=0.01\Gamma_3$ (blues lines), respectively. Other parameters are $\Delta p=\Delta= g\sqrt{N} =20\Gamma_3$, $\Omega=0.2\Gamma_3$ and $\kappa=3\Gamma_3$,

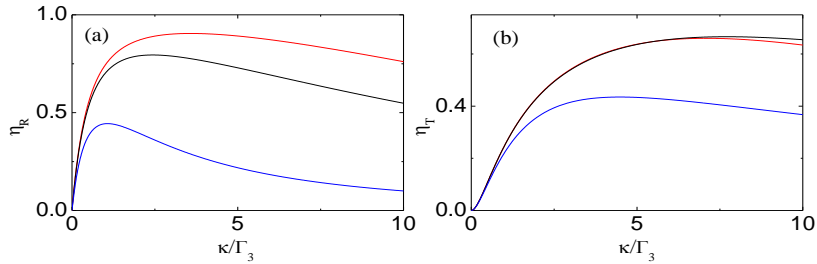

Fig. 6 (a) Switching efficiency of the reflected signal light $\eta_R$ and (b) switching efficiency of the transmitted signal light $\eta_T$ versus $\kappa/\Gamma_3$ with $\gamma_{12}=0.0001\Gamma_3$ (red lines), $\gamma_{12}=0.001\Gamma_3$ (black lines) and $\gamma_{12}=0.01\Gamma_3$ (blues lines), respectively. Other parameters are $\Delta p=\Delta= g\sqrt{N} =20\Gamma_3$, and $\Omega=0.2\Gamma_3$.

The switching efficiency $\eta_R$ and $\eta_T$ depend on the collective coupling coefficient $g\sqrt{N}$, the control laser Rabi frequency $2\Omega$, the cavity decay rate $\kappa$, and the deocherence rate $\gamma_{21}$. In order to clarity the performance characteristics of the coupled cavity-atom system, we plot in Fig. 3 to Fig. 6 separately the switching efficiency $\eta_R$ and $\eta_T$ versus these parameters with practical parameter values obtainable experimentally [22]. Fig. 3 shows that $\eta_R$ and $\eta_T$ are monotonically increasing functions of $g\sqrt{N}$ and are saturated at moderate $g\sqrt{N}$ value ($g\sqrt{N} \geq 15\Gamma_3$). Fig. 4 shows that $\eta_R$ and $\eta_T$ decreases monotonically with increasing $\gamma_{21}$, which indicates that the interaction-free all-optical switching is a coherent process and the switching efficiency

(particularly $\eta_R$) depends sensitively on $\gamma_{21}$. For the high efficiency operation of the all-optical switching, it is desirable to have an atomic system with small decay rate of the ground state coherence. It has been shown [36] that in cold Rb atoms, the decoherence rate as small as $\gamma_{12}=10^{-4}\Gamma_3$ has been observed. Therefore, it is possible to achieve high switching efficiencies in experiments with cold alkaline atoms as the optical medium.

Fig. 5 shows that the interaction-free all-optical switching in the cavity-atom system can be done with a weak control laser. The switching efficiency increases rapidly with the increasing control field, but saturates at $\Omega$ values smaller than $\Gamma_3$, i.e., below the saturation intensity of the control field transition. Fig. 6 plots $\eta_R$ and $\eta_T$ versus the cavity decay rate $\kappa$ and shows that the switching efficiency can be optimized at a relatively high $\kappa$ value ($<5\Gamma_3$ for the parameters used in Fig. 6). Overall, the highly efficient all-optical switching at low control intensities requires a moderately large collective coupling coefficient $g\sqrt{N}$, a sufficiently large cavity decay rate $\kappa$, and a small decoherence rate $\gamma_{21}$. These requirements can be readily fulfilled experimentally. As a numerical example, consider cold Rb atoms ($\Gamma_3$=3 MHz) confined in a 5 cm cavity with a finesse of 150 ($\kappa$=10 MHz), with $g\sqrt{N}$ =50 MHz (N≈$10^4$ atoms), $\gamma_{21}$=10 KHz(~0.003$\Gamma_3$), and $\Omega$=1.5 MHz (corresponding to a control intensity $I_c = c\varepsilon_0 E^2 = c\varepsilon_0(\hbar\Omega/\mu_{23})^2 \approx 0.3$ mW/cm$^2$ that is about 5 times smaller than the Rb saturation intensity of 1.6 mW/cm$^2$), the switching efficiency is derived to be $\eta_R$=0.83 and $\eta_T$=0.56. Fig. 3-6 also show that the switching efficiencies for the two output channels are different. Under the normal operating conditions discussed here, $\eta_R > \eta_T$: it is more efficient to operate the all-optical switch of the cavity-atom system in the reflection mode.

## 4. Switching time and frequency bandwidth

The minimum switching time is an important parameter for all-optical switching applications. It is desirable to have an optical switch operating with a fast switching time, and at the same time, low control intensity. A fast switching time requires a large frequency bandwidth for the optical switching system, but it is usually incompatible with the low intensity requirement. The above analysis shows that the interaction-free all-optical switching in the multi-atom cavity QED system is capable of operating at low intensities. We next discuss the switching bandwidth and switching time and their relationship with the switching intensity for the cavity-atom system. As shown in Fig. 2, the control laser induces the destructive interference in the normal mode excitation and the transmission (reflection) spectrum of the signal light intensity versus the signal frequency detuning Δp exhibits a dip (peak) across the resonance of the normal mode transition. The linewidth of the dip (peak) gives the frequency bandwidth of the system and determines the minimum switching time. At the normal mode resonance ($\Delta_p = g\sqrt{N}$), the intensity of the transmission channel $I_T(\Omega=0)$ is at the maximum when the control laser is absent ($\Omega=0$); when the control laser is present ($\Omega\neq 0$), the intensity of the transmission channel $I_T(\Omega\neq 0)$ is at the minimum (dip). Then, let the signal laser frequency detuning be $\Delta_p = g\sqrt{N} \pm \delta$, at which the intensity of the transmission channel becomes $I_T(\delta) = \frac{1}{2}(I_T(\Omega=0,\delta=0) + I_T(\Omega,\delta=0))$, the half-width δ at the half dip minimum (HWHM) can be derived. Under the condition $g\sqrt{N} \gg \kappa$, Ω, and $\Gamma_3$, and $\gamma_{21} \ll \kappa$ and $\Gamma_3$, it is given by

$$\delta = \frac{\sqrt{(\Gamma_3+\kappa)^2 + 8\Omega^2} - \kappa - \Gamma_3}{4}. \quad (10)$$

2δ is the frequency bandwidth of the all-optical switching system and then the minimum switching time is $\tau = \frac{1}{4\pi\delta} = \frac{1}{\pi(\sqrt{(\Gamma_3+\kappa)^2 + 8\Omega^2} - \kappa - \Gamma_3)}$, in which $\Omega^2 = \frac{\mu_{23}^2 I_c}{h^2 c\varepsilon_0}$ ($I_c$ is the control light intensity). The switching frequency bandwidth δ can also be derived by considering the peak of the reflected signal light near $\Delta_p = g\sqrt{N}$. With $I_R(\delta) = \frac{1}{2}(I_R(\Omega=0,\delta=0) + I_R(\Omega,\delta=0))$, the identical equation for δ as Eq. (10) is derived under the condition $g\sqrt{N} \gg \kappa$, Ω, and $\Gamma_3$, and $\gamma_{21} \ll \kappa$ and $\Gamma_3$. Eq. (10) shows that the switching frequency bandwidth increases with the increasing control light intensity. Therefore shorter switching times require higher control intensities. Thus, the lower intensity operation of the interaction-free all-optical switching in the cavity-atom system is accompanied by the increased switching time. If we consider $\Omega^2 \ll (\kappa+\Gamma_3)$, the minimum switching time becomes

$$\tau \approx \frac{\kappa+\Gamma_3}{4\pi\Omega^2} = \frac{h^2 c\varepsilon_0(\kappa+\Gamma_3)}{4\pi\mu_{23}^2 I_c},$$

which is inversely proportional to the control intensity $I_c$. As a numerical example for a practical cavity-atom system (with cold $^{85}$Rb atoms and a 5 cm near cavity [22,27], $\Omega=3$ MHz, $g\sqrt{N}=50$ MHz, $\Gamma_3=3$ MHz, $\gamma_{21}=0.01$ MHz, and $\kappa=20$ MHz), the switching efficiency for the reflection channel is $\eta_R\approx0.98$, the switching efficiency for the transmission channel is $\eta_T\approx0.75$, and the minimum switching time is $\tau\approx0.2$ μs.

Next we analyze the propagation of a Gaussian input signal pulse, $a_p^{in}(t)=E_0\exp(-t^2/2\sigma^2)$ ($\sigma$ is the pulse standard deviation) through the cavity-atom system and derive the cavity reflected signal pulse and the cavity transmitted signal pulse. Assuming the control field pulse is sufficiently long such that it overlaps with the intra-cavity signal pulse completely and can be viewed as having constant amplitude for the duration of the signal pulse. Substituting the Fourier transformation for $\sigma_{ij}(t)$ (i,j=1-3) and $a(t)$,

$$\sigma_{ij}(t)=\sqrt{2\pi}\int_{-\infty}^{\infty}\sigma_{ij}(\omega)\exp(i\omega t)d\upsilon, \text{ (i,j=1-3)}, \qquad (11)$$

$$a(t)=\sqrt{2\pi}\int_{-\infty}^{\infty}a(\omega)\exp(i\omega t)d\upsilon, \qquad (12)$$

into equations (2) and a set of equations for $\sigma_{ij}(\omega)$ and $a(\omega)$ are derived. Consider the condition of $g\ll\Omega$ such that $\sigma_{11}=1$, the Fourier transformation of the intra-cavity field $a$ can be derived as

$$a(\omega)=\frac{\sqrt{\kappa/\tau}a_p^{in}(\omega)}{\kappa-i(\omega+\Delta_c-\Delta_p)+\dfrac{g^2N}{\Gamma_3-i(\omega+\Delta_p)+\dfrac{\Omega^2}{\gamma_{12}-i(\omega+\Delta_p-\Delta)}}}. \qquad (13)$$

Here the Fourier transform of the input signal field is $a_p^{in}(\omega)=\sqrt{2\pi}\sigma E_0\exp(-\sigma^2(\omega-\omega_p)^2/2)$. The Fourier transforms of the transmitted signal field is $a_T(\omega)=\sqrt{\kappa\tau}a(\omega)$ and the reflected signal field is $a_R(\omega)=\sqrt{\kappa\tau}a(\omega)-a_p^{in}(\omega)$, respectively. Then, the transmitted signal pulse and the reflected signal pulse are given by $a_T(t)=\sqrt{2\pi}\int_{-\infty}^{\infty}a_T(\omega)\exp(i\omega t)d\upsilon$ and $a_R(t)=\sqrt{2\pi}\int_{-\infty}^{\infty}a_R(\omega)\exp(i\omega t)d\upsilon$, respectively. The intensity of the transmitted signal pulse and the reflected signal pulse are then $I_T(t) = a_T(t)a_T^*(t)$ and $I_R(t) = a_R(t)a_R^*(t)$, respectively. Fig. 7 and Fig. 8 plot the reflected signal pulses and the transmitted signal pulses with the control field ($\Omega=0.5\Gamma_3$, blue lines) and without the control field ($\Omega=0$, red lines) for 3 different time durations of the input signal pulse normalized with a unity peak intensity, respectively. With $\Omega=0.5\Gamma_3$ for the control field and the given parameters in Fig. 7 and 8, the switching frequency bandwidth $2\delta=0.12\Gamma_3$, and the corresponding switching time is $\tau=1/(4\pi\delta)=1.3/\Gamma_3$. If the input signal pulse has a duration T>τ, then there is minimal pulse distortion and the high efficiency switching for the signal pulse can be realized as shown by Fig. 7(a) and 8(a). As the input signal pulse duration decreases, the output signal pulse becomes distorted and the switching efficiency decreases as shown in Fig. 7(b) (8(b)) and 7(c) (8(c)). The pulse distortion occurs only when the control laser is present ($\Omega\neq0$). When the control laser is not present, the transmitted signal light is at the maximum and the reflected signal is at the minimum, and there is no observable pulse distortion. This can be understood as follows: when the control light is not present ($\Omega=0$), the linewidth of the transmission spectral peak (the reflection valley) shown in Fig. 2(b) (Fig. 2(a)) is given by $(\kappa+\Gamma_3)/2$ [35], which is larger than the frequency bandwidth of the input pulses used in Fig. 7 and 8. Therefore there is no pulse distortion for the transmitted signal pulse and the reflected signal pulse. When the control field is present ($\Omega\neq0$) and induces the quantum interference that suppresses the normal mode excitation, the spectral linewidth of the induced interference dip for the transmitted signal light (peak for the reflected signal light) is given by Eq. (10), which is smaller than the frequency bandwidth of the input signal pulse (with the pulse duration $\sigma=3/\Gamma_3$ in (b) and $\sigma=1.5/\Gamma_3$ in (c)). Therefore, the pulse distortion is observed for the output signal pulse in both the transmission channel and the reflection channel in Fig. 7(b) and 7(c) (Fig. 8(b) and 8(c)), The switching bandwidth also limits the switching efficiency for the input signal pulse with a large frequency bandwidth (a shorter pulse duration). When this happens, the amplitude of the reflected signal pulse decreases while the amplitude of the transmitted signal pulse increases, indicating that when the transmission channel is in the open state, the light leakage increases with the decreasing pulse duration.

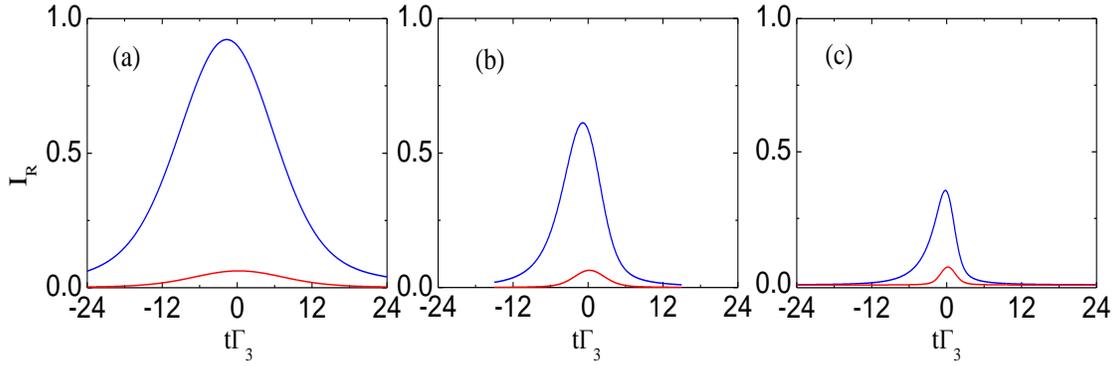

Fig. 7 Reflected signal pulse versus time $t\Gamma_3$. $\Omega=0.5\Gamma_3$ in blues lines and $\Omega=0$ in the red lines. In (a), $\sigma=9/\Gamma_3$; in (b), $\sigma=3/\Gamma_3$; and in (c), $\sigma=1.5/\Gamma_3$. The other parameters are $g\sqrt{N}=20\Gamma_3, \kappa=3\Gamma_3$, $\Delta=\Delta p= g\sqrt{N}=20\Gamma_3$, $\Delta_c=\Delta_p=0$ and $\gamma_{12}=0.0001\Gamma_3$.

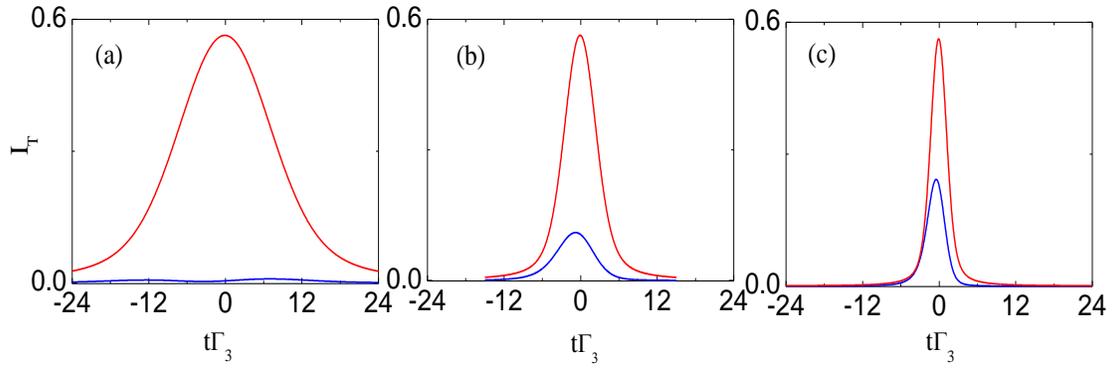

Fig. 8 Transmitted signal pulse versus time. $\Omega=0.5\Gamma_3$ in blues lines and $\Omega=0$ in the red lines. The parameters are the same as those of Fig. 7.

## 5. Conclusion

We have shown that interaction-free all-optical switching can be realized in a coherently coupled multi-atom cavity QED system. The system supplies two complementary output channels for the input signal light controlled by a single light field coupled to the intra-cavity atoms from free pace. The analytical results and numerical calculations show that the free-space control light induces quantum interference that suppresses the normal mode excitation and enables the all-optical switching at low light intensities. The switching frequency bandwidth of the cavity-atom system is determined by the linewidth of the interference-induced spectral peak (dip) of the cavity reflected (transmitted) signal light and increases with the increasing control light intensity. A faster switching time necessitates an increase of the control light intensity or vice versa. High efficiency all-optical switching, particularly for the reflection channel operation, can be obtained under practical conditions. The multi-atom cavity QED system may be also used to explore the cross-phase modulation of signal light by the control light and the two output channels from the reflection and transmission offers versatility and flexibility, which may be useful for quantum electronics and photonics applications.

References

1. S. E. Harris and Y. Yamamoto, Phys. Rev. Lett. 81, 3611(1998).
2. B. S. Ham and P. R. Hemmer, Phys. Rev. Lett. 84, 4080(2000).
3. K. J. Resch , J. S. Lundeen, A. M. Steinberg, Phys. Rev. Lett. 89, 037904(2002).
4. H. Schmidt and R. J. Ram, Appl. Phys. Lett. 76, 3173(2000).
5. M. Yan, E. Rickey, and Y. Zhu, Phys. Rev. A **64**, 041801(R) (2001).
6. D. A. Braje, V. Balić, G. Y. Yin, and S. E. Harris, Phys. Rev. A 68, 041801(R) (2003).
7. A. Dawes, L. Illing, S. M. Clark, D. J. Gauthier, Science 29, 672(2005).
8. H. Mabuchi, Phys. Rev. A 80, 045802 (2009).
9. D. Chang, A. Sørensen, E. Demler, and M. Lukin, Nat. Phys. 3, 807 (2007).
10. M. Albert, A Dantan, and M. Drewsen, Nature Photonics 5, 633 (2011).
11. Jiteng Sheng, Utsab Khadka, and Min Xiao, Phys. Rev. Lett. 109, 223906 (2012).
12. Y. F. Chen, Z. H. Tsai, Y. C. Liu, and I. A. Yu, Opt. Lett. 30, 3207(2005).
13. H. Wang, D. Goorskey, and M. Xiao, Opt. Lett. **27**, 1354 (2002).
14. M. Soliacic, E. Lidorikis, J. D. Joannopoulos, L. V. Haus, Appl. Phys. Lett. 86, 171101(2005).
15. C. Y. Wang, Y. F. Chen, S. C. Lin, W. H. Lin, P. C. Kuan, and I. A. Yu, Opt. Lett. 31, 2350(2006).
16. J. Zhang, G. Hernandez, and Y. Zhu, Optics Express **16**, 19112(2008); J. Zhang, G. Hernandez, and Y. Zhu, Opt. Lett. **32**, 1317(2007).
17. B. C. Jacobs and J. D. Franson, Phy. Rev. A 79, 063830(2009) (all-optical switching-quantu Zeno 2-photon absorption).
18. Y. Huang and P.Kumar, Opt. Lett. 35, 2376(2010).
19. Y. Huang, J. B. Altepeter, and P. Kumar, Phys. Rev. A 82, 063826(2010).
20. K. T. McCusker, Y. Huang, A. S. Kowligy, and P. Kumar, Phys. Rev. Lett. 110, 240403(2013). X2 based (17 W peak power).
21. S. M. Hendrickson, C. N. Weiler, R. M. Camacho, P. T. Rakich, A. I. Young, M. J. Shaw, T. B. Pittman, J. D. Franson, and B. C. Jacobs, Phy. Rev. A 87, 023808 (2013).
22. J. Zhang, G. Hernandez, and Y. Zhu, Optics Express **16**, 7860(2008).
23. Y. Zhu, Opt. Lett. **35**, 303(2010).
24. X. Wei, J. Zhang, Y. Zhu, Phys. Rev. A 82, 033808 (2010).

25. M. D. Lukin, M. Fleischhauer, M. O. Scully, and V. L. Velichansky, Opt. Lett. **23**, 295-297 (1998).

26. H. Wang, D. J. Goorskey, W. H. Burkett, and M. Xiao, Opt. Lett. **25**, 1732-1734 (2000).

27. G. Hernandez, J. Zhang, and Y. Zhu, Phys. Rev. A **76**, 053814 (2007).

28. H. Wu, J. Gea-Banacloche, and M. Xiao, Phys. Rev. Lett. **100**, 173602 (2008).

29. A. E. B. Nielsen, J. Kerckhoff, Phys. Rev. A 84, 043821 (2011).

30. B. Zou and Y. Zhu, Phys. Rev. A **87**, 053802(2013).

31. A. Dantan, M. Albert, and M. Drewsen, Phys. Rev. A 85, 013840(2012).

32. D. F. Walls and G. J. Milburn, "*Quantum Optics*", (Springer-Verlag, Berlin, Heidelberg, 1994).

33. J. J. Sanchez-Mondragon, N. B. Narozhny, and J. H. Eberly, Phys. Rev. Lett. 51, 550(1983).

34. G. S. Agarwal, Phys. Rev. Lett. 53, 1732(1984).

35. A. Boca, R. Miller, K. M. Birnbaum, A. D. Boozer, J. McKeever, and H. J. Kimble, Phys. Rev. Lett. 93, 233603 (2004).

36. R. Zhao, Y. O. Oudin, S. D. Jenkins, C. J. Campbell, D. N. Matsukevich, T. A. B. Kenndedy, and A. Kuzmich, Nature Phys. **5**, 100(2009).